\documentclass[aps,prc,nofootinbib,showpacs,preprint]{revtex4}

\usepackage{epsfig}
\usepackage{graphicx}
\usepackage[centertags]{amsmath}

\newcommand\ba{\begin{eqnarray}}
\newcommand\ea{\end{eqnarray}}

\newcommand{\be}{\begin{equation}}
\newcommand{\ee}{\end{equation}}

\begin{document}

\title{Two-component model for the axial form factor of the nucleon}

\author{C. Adamu\v s\v c\'in }
\altaffiliation{\it Department of Theoretical Physics, IOP, Slovak Academy of Sciences, Bratislava, Slovakia}
\affiliation{\it CEA,IRFU,SPHN, Centre de Saclay, F-91191 Gif-sur-Yvette, France}

\author{E. Tomasi-Gustafsson}
\email{etomasi@cea.fr}
\affiliation{\it CEA,IRFU,SPHN, Centre de Saclay, F-91191 Gif-sur-Yvette, France}

\author{E. Santopinto}
\affiliation{\it I.N.F.N., via Dodecaneso 33,
Genova, I-16146, Italy}

\author{R. Bijker}
\affiliation{\it Instituto de Ciencias Nucleares, 
Universidad Nacional Aut\'onoma de M\'exico, AP 70-543, 04510 Mexico DF, Mexico}

\begin{abstract}
The axial form factor of the nucleon is studied in a two-component model consisting of a 
three-quark intrinsic structure surrounded by a meson cloud. The experimental data in the 
space-like region are well reproduced with a minimal number of parameters. The results are 
similar to those obtained from a dipole fit for $0 < Q^2 < 1$ GeV$^2$, but outside this 
region there are important deviations from the dipole parametrization. Finally, the 
theoretical formula for the axial form factor is extrapolated by analytic continuation to 
the time-like region, thus providing the first predictions in this kinematical region which 
is of interest for present and future colliders. 
\end{abstract}


\maketitle

\section{Introduction}

The electroweak structure of the nucleon is characterized by both electromagnetic 
and weak form factors and, in particular, by the weak axial form factor, $G_A(Q^2)$ ($Q^2$ is the four momentum transfer squared), 
which is related to the nucleon axial current. The existing experimental information 
on the axial form factor in the space-like region can be obtained directly through the 
reaction $\nu_{\mu}+p\to \mu^++n$, or indirectly through charged pion electroproduction 
near-threshold experiments \cite{Be02}. Axial form factors also play an important role 
in the analysis of parity violating electron scattering. Especially, in order to  
extract information on the strange form factors of the proton requires a good knowledge 
of the axial form factor \cite{pves}. 

Predictions on the axial for factor have been given in different models which describe 
the nucleon structure, such as the chiral constituent quark model \cite{Gl01,Ba06}, 
the chiral perturbation theory \cite{Sc06}, the quark-soliton model \cite{Si05}, the light 
cone QCD sum-rules \cite{Wa06}. Results from lattice QCD have become available recently  
\cite{Al07}. 

The axial form factor is usually parametrized by means of a dipole form \cite{Be02}   
which gives a reasonable description of the data up to $Q^2=1$ GeV$^2$ covering  the range of most of the available measurements. It is useful to have other parametrizations 
\cite{Bo08}, even though it is difficult to discriminate among them on the basis of the 
existing data alone. Indeed for many years, the dipole parametrization was considered to 
provide a very good description of the proton and neutron magnetic form factors and the 
electric proton form factor, whereas the electric neutron form factor was assumed to be 
zero or very small and well described, for example, by the Galster parametrization 
\cite{Ga71}. However, it has recently been shown that the electric and magnetic form 
factors of the proton are actually very different, and that the ratio $\mu G_E^p/G_M^p$ 
drops almost linearly as a function of $Q^2$ \cite{Jo00}, in contrast with the dipole 
description. 

The Iachello, Jackson and Land\'e model (IJL) \cite{Ia73} predicted this behavior 
by means of a two-component model for the electric proton form factor long before 
the data appeared. More recently, Bijker and Iachello (BI) \cite{Bi04} have shown 
that it is possible to refine the two-component model in oder to reproduce further 
details, in particular, concerning the electric and magnetic form factors of the neutron. 
The IJL and BI approaches are based on a two-component picture of the nucleon 
in terms of an intrinsic structure ($qqq$ configuration) surrounded by a meson cloud 
($q\bar{q}$ pairs). It has been shown to be rather successful in the description of 
the nucleon electromagnetic form factors both in the space- and in the
time-like 
region \cite{Bi04,Wa05,Ia04}. Other applications of the two-component model include the deuteron 
\cite{ETG06} and the strange form factors of the proton \cite{jpg}.

The purpose of this paper is to apply the two-component model of nucleon form factors 
\cite{Ia73,Bi04} to the axial form factor, and to study its analytic continuation to 
the time-like region for which the axial form factor has not yet been measured. 
Suggestions for its determination through the reaction $N \bar p \to \gamma^* N \pi$ 
and the crossed channels can be found in \cite{Re65,Du95,Ad07}. 
This problem can become very actual in connection with the physics planned with the 
antiproton beam which will be available at the FAIR accelerator complex. 

\section{Axial form factors}

The axial form factor has been measured directly in neutrino scattering, 
$\nu_{\mu}+p\to \mu^++n$, or indirectly, in near-threshold charged pion electroproduction 
in the space-like region. In both reactions the axial form factor is linked to weak charged 
currents. The available experimental information is usually parametrized in terms of a 
dipole \cite{Be02}
\begin{equation} 
G_A^D(Q^2) = \frac{G_A(0)}{(1+Q^2/M_A^2)^2} ~.
\label{eq:dipole}
\end{equation}
At $Q^2=0$, the axial form factor can be determined from neutron $\beta$ decay as 
$G_A(0)=1.2695 \pm 0.0029$ \cite{pdg2006}. 
The axial mass $M_A$ is adjusted to the experimental data. From charged pion 
electroproduction one obtains $M_A=1.069\pm 0.018 $ GeV, whereas in neutrino scattering 
experiments, $M_A$ is extracted from a weighted average to be $M_A=1.026\pm 0.021$ GeV, 
which is somehow inconsistent with the best fit value obtained from the electroproduction 
experiments. Even if the neutrino data suffer from great uncertainties, the weighted average 
for the root mean square radius and thus also for $M_A$ ($\langle r^2 \rangle_A=12/M_A^2$) 
is considered to be quite reliable. 

Similarly to the Rosenbluth separation for electromagnetic form factors, the axial 
(pseudoscalar) form factor is related to the slope (intercept) of the near threshold 
differential cross section as a function of the polarization of the virtual photon. 
By means of low energy theorems it is possible to calculate the electric dipole 
amplitude at the threshold in the case of soft pions. Model-dependent corrections, 
have to be introduced in order to take into account the finite pion mass. 
It has been shown in \cite{Be02} that if one takes into account the corrections due to 
the finite pion mass in chiral perturbation theory which should be applied to the root 
mean square axial radius as extracted from charged pion electroproduction data, they do 
indeed correspond to an increase in the root mean square value. This leads to a lowering 
of the $M_A$ value as extracted from electroproduction of the order of 5\%, which 
makes it compatible with the neutrino value. 

Additional experimental information on the axial form factor may be obtained from 
weak neutral current processes in parity violating electron scattering experiments.  
There is a proposal of the G0 collaboration for dedicated runs at backward angles 
in order to extract information on the axial coupling of the photon with the nucleon 
\cite{G0}. The SAMPLE experiment yielded values for the axial form factor by combining  
the results for proton and deuteron targets \cite{Sample}.  

\section{Two-component model}

In the two-component model \cite{Ia73,Bi04}, the axial nucleon form factor is 
described as 
\begin{eqnarray}
G_A(Q^2) &=& G_A(0) \, g(Q^2) \left[ 1-\alpha +\alpha \frac{m_A^2}{m_A^2+Q^2} \right] ~, 
\nonumber\\
g(Q^2) &=& \left (1+\gamma Q^2\right )^{-2} ~,
\label{eq:eq1}
\end{eqnarray}
with $Q^2>0$ in the space-like region. $g(Q^2)$ denotes the coupling to the intrinsic 
structure (three valence quarks) of the nucleon, and $m_A$ is the mass of the lowest 
axial meson $a_1(1260)$ with quantum numbers $I^G(J^{PC})=1^-(1^{++})$ and $m_A= 1.230$ GeV.  
We note that, unlike other studies, in which  $m_A$ is a parameter, here 
it corresponds to the mass of the axial meson $a_1(1260)$. In the present case,  
$\gamma$ is taken from previous studies of the electromagnetic form factors of the 
nucleon \cite{Ia73,Bi04}. Therefore, $\alpha$ is the only fitting parameter, 

It is interesting to note, that this form of the axial form factor can give rise 
to a zero in the space-like region. If $\alpha >1$, the axial form factor goes through 
zero at $Q^2=m_A^2/(\alpha-1)$. Since for large values of $Q^2$ the contribution 
of the axial meson cloud vanishes, the asymptotic behavior of the axial form factor of  
Eq.~(\ref{eq:eq1}) is given by its intrinsic part only
\begin{equation}
\lim_{Q^2 \rightarrow \infty} G_A(Q^2) = \frac{G_A(0)(1-\alpha)}{(\gamma Q^2)^2} ~,
\label{eq:eqas}
\end{equation}
which becomes negative if $\alpha >1$.

The behavior of the axial form factor at low values of $Q^2$ can be used to 
determine the axial radius 
\ba
\langle r^2 \rangle_A &=& -6 \left.\frac{d G_A(Q^2)}{d Q^2}\right|_{Q^2=0} 
= \left\{ \begin{array}{cl} \frac{12}{M_A^2} & \hspace{1cm} \mbox{dipole} \\ 
6\left(2\gamma+\frac{\alpha}{m_A^2}\right) 
& \hspace{1cm} \mbox{two-component} \end{array} \right.
\label{radius}
\ea
A comparison of the axial radius for the dipole and the two-component model 
may be used to express the coefficient $\alpha$
\be
\alpha = 2m_A^2\left (\displaystyle\frac{1}{M_A^2}-\gamma \right ) ~,
\label{alpha}
\ee  
in terms of the mass of the lightest axial meson $m_A$, the fitted value of 
the axial mass $M_A$ appearing in the dipole form and $\gamma$, which is 
proportional to the intrinsic radius. 

\section{Analysis of the data in the space-like region}

In this section, we study the axial form factor of the proton in a two-component model. 
The experimental data are taken from a compilation of pion electroproduction experiments 
on the nucleon \cite{Be02}. Since the present neutrino data suffer severe uncertainties 
\cite{Be02}, in the present analysis we only consider the pion electroproduction data. 

\begin{figure}
\mbox{\epsfxsize=14.cm\leavevmode \epsffile{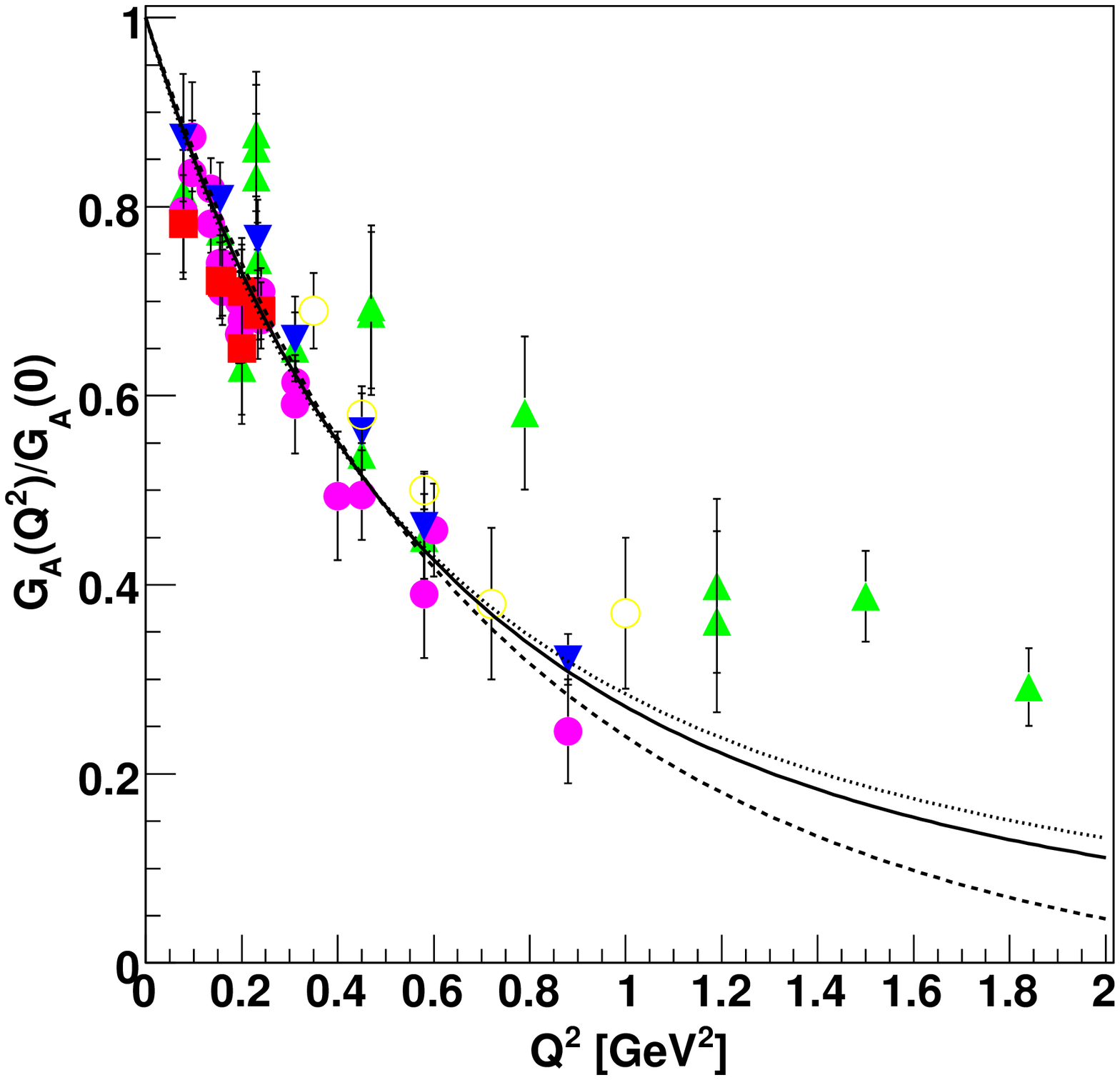}}
\vspace*{.2 truecm}
\caption{(Color online) Comparison between theoretical and experimental values of the 
axial form factor of the nucleon $G_A(Q^2)$ as a function of $Q^2$.  
The theoretical values are calculated in the two-component model using Eq.~(\ref{time}) with $\alpha=1.57$ and $\gamma = 0.25$ GeV$^{-2}$ \cite{Ia73} (dashed line), 
and $\alpha = 0.95$ and $\gamma = 0.515$ GeV$^{-2}$ \cite{Bi04} (solid line), 
and the dipole form of Eq.~(\ref{eq:dipole}) with $M_A = 1.069$ GeV (dotted line). 
The experimental values were extracted according different models: PCAC \protect\cite{Be73} 
(pink, solid circles), FPV (red, solid squares) \protect\cite{Fu70},  
SP (green, solid triangles) \protect\cite{Na70}, DR  (blue, trianglesdown)  
\protect\cite{Do73}, $\Delta$ (yellow, open circles) \protect\cite{Jo76}.}
\label{fig:axfit}
\end{figure}

The $Q^2$ dependence of the nucleon axial form factor $G_A(Q^2)$, has been measured in 
several pion electroproduction experiments at threshold over the last few decades. 
The slope of the total unpolarized differential cross section at threshold contains 
information on $G_A(Q^2)$, but the numerical value of this form factor is highly 
model-dependent. In general, four different approaches have been used to extract 
the values of the axial form factor of the nucleon: the Soft Pion approximation (SP) 
\cite{Na70}, the Partially Conserved Axial Current approximation (PCAC) \cite{Be73}, 
the  Furlan approximation (FPV) \cite{Fu70} (enhanced soft pion production) and the 
Dombey and Read approximation (DR) \cite{Do73}. As a consequence of these competing 
approaches, up to four experimental values may be extracted from a single measurement 
(at fixed $Q^2$). A total of 67 experimental points are available, corresponding to 
32 measurements. Data from Ref.~\cite{Jo76} were considered separately, as they 
correspond to $\Delta$ excitation in the final state. In order to evaluate 
the systematic error, the data were therefore separated into 5 groups according to the 
approach used and the processes measured. The data from \cite{Bl73} 
were not considered in the fit, following Ref.~\cite{Be02}, 
as they are systematically larger, nor were data from \cite{Na70}.
The data, normalized to one, are plotted in Fig.~\ref{fig:axfit}. 
Different symbols correspond to different models used for the extraction 
of the data but may correspond to the same experiment. 

The form factor $g(Q^2)$ in the two-component model describes the coupling to the 
intrinsic structure (three valence quarks) of the nucleon, where $\gamma$ was 
determined from a fit of nucleon electromagnetic form factors to be $\gamma=0.25$ 
GeV$^{-2}$ \cite{Ia73} or $\gamma=0.515$ GeV$^{-2}$ in a more recent fit \cite{Bi04}.  
We note however that the former value is not good from a $t$ channel point of 
view, because it gives a pole in the physical region at $t_0=1/\gamma=4$ GeV$^{2}$ 
($>4m^2=3.52$ GeV$^{2}$, the corresponding threshold). In the latter case, the pole 
is shifted to the unphysical region. In our calculations of the axial form factors 
we keep $\gamma$ as a fixed parameter, and consider both values mentioned above.  

Individual one--parameter fits to the 5 data sets were performed, as well as a global 
fit, according to Eq.~(\ref{eq:eq1}). The results are shown in Table~\ref{tab1} 
and in Fig.~\ref{fig:axfit}. The global fit gives $\alpha = 1.57 \pm 0.04$ with 
$\chi^2/{\rm n.d.f.} = 85.36/48 = 1.78$ for $\gamma = 0.25$ GeV$^{-2}$  
\cite{Ia73}, and $\alpha = 0.95 \pm 0.05$ with 
$\chi^2/{\rm n.d.f.} = 69.60/48 = 1.45$ for $\gamma = 0.515$ GeV$^{-2}$ \cite{Bi04}. 
In both cases, the $\chi^2$ for individual fits may be smaller 
than the global $\chi^2$, owing to the dispersion of the data, but the errors associated 
to the parameters of the global fits are smaller, owing to the larger number of points. 
These values of $\alpha$ can be considered as an average of the different corrections. 
The associated systematic error, which takes into account the dispersion of the model 
analysis, can be evaluated from the results of the individual fits to be $<|0.35|$. 

\begin{figure}
\mbox{\epsfxsize=14.cm\leavevmode \epsffile{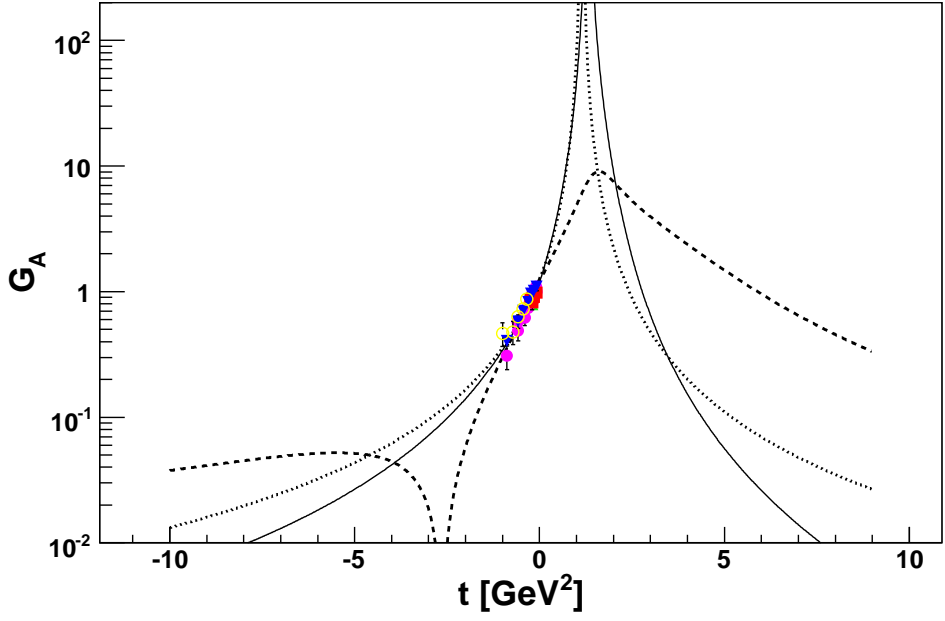}}
\vspace*{.2 truecm}
\caption{(Color online) As Fig.~\ref{fig:axfit}, but for the absolute value of the 
axial form factor $|G_A(t)|$ in the space-like ($t<0$) and time-like ($t>0$) regions. 
In the time-like region, $\delta=0.925$ for IJL \cite{Wa05} and $\delta=0.397$ for BI 
\cite{Bi04}.}
\label{fig:2}
\end{figure}

In Fig.~\ref{fig:axfit} we show a comparison between the experimental and theoretical 
values of the axial form factor for the dipole fit, the global fit with $\alpha=1.57$ 
and $\gamma = 0.25$ GeV$^{-2}$ \cite{Ia73}, and $\alpha = 0.95$ and 
$\gamma = 0.515$ GeV$^{-2}$ \cite{Bi04}.  
It is possible to give a reasonable description of the data, if we consider the average 
value, since we average not only statistical errors, but also the systematic errors 
related to the model-dependence extraction of the data. 
In the range up to $Q^2=1$ GeV$^2$ the description of the data is comparable to the 
quality of a dipole fit, though  it is clear that already around $Q^2=1$ GeV$^2$ the 
three parametrizations start to show a different behavior.
According to Eq.~(\ref{eq:eqas}), the two fitted values of the $\alpha$ parameter imply 
a different asymptotic behavior with a change of sign at $Q^2=2.65$ GeV$^2$ for the IJL 
parametrization \cite{Ia73}, but not for BI \cite{Bi04}, nor for the dipole (see also  
Fig.~\ref{fig:2}). 

\begin{table}[t]
\centering
\begin{tabular}{|c|c|c|c|c|c|c|c|}
\hline
Model & DR & FPV & SP & PCAC & $\Delta$ & Global & Ref. \\
\hline
$\alpha$ & $1.38 \pm 0.08$ & $1.90 \pm 0.14$ & $1.15 \pm 0.06$ 
& $1.80 \pm 0.06$ & $1.21 \pm 0.07$ & $1.57 \pm 0.04$ & \protect\cite{Ia73}\\
$\chi^2/{\rm n.d.f.}$ & $0.19$ & $0.81$ & $3.9$ & $0.79$ & $0.43$ & $1.78$ & \\
\hline 
$\alpha$ & $0.75 \pm 0.10$ & $1.23 \pm 0.15$ & $0.52 \pm 0.09$ 
& $1.17 \pm 0.07$ & $0.53 \pm 0.09$ & $0.95 \pm 0.05$ & \protect\cite{Bi04}\\
$\chi^2/{\rm n.d.f.}$ & $0.40$ & $0.75$ & $3.45$ & $0.67$ & $0.49$ & $1.45$ & \\
\hline  
\end{tabular}
\caption[]{Fitted $\alpha$ parameter and corresponding $\chi^2/{\rm n.d.f.}$ for the different model-dependent extractions of the axial data.}
\label{tab1}
\end{table}

The values of $\alpha$ obtained in the global fits are close to the values that can be 
derived from Eq.~(\ref{alpha}) with $m_A = 1.230$ GeV and $M_A = 1.069$ GeV 
in which it is assumed that the axial radius for the dipole and the two-component 
model is the same: $\alpha = 1.89$ for $\gamma = 0.25$ GeV$^{-2}$ \cite{Ia73} 
and $\alpha = 1.09$ for $\gamma = 0.515$ GeV$^{-2}$ \cite{Bi04}. 

The axial radius $\sqrt{\langle r^2 \rangle_A}$ can be obtained from Eq.~(\ref{radius}): 
0.60 fm for IJL, 0.62 fm for BI and 0.64 fm for the dipole. In the two-component model 
the contributions of the quark core and the axial meson cloud to the axial radius 
are given by  
\begin{eqnarray}
\langle r^2 \rangle_A = \left\{ \begin{array}{cl} 
12 \gamma (1-\alpha) & \hspace{1cm} \mbox{quark core} \\ 
6\alpha\left(2\gamma+\frac{1}{m_A^2}\right) & \hspace{1cm} \mbox{axial meson cloud} 
\end{array} \right.
\label{intaxm}
\end{eqnarray}
The difference between the two parametrizations of the two-component model 
(IJL and BI) is in the values of $\gamma$ and $\alpha$. The value of $\gamma$ 
corresponds to the spatial extent of the intrinsic dipole form factor 
$\langle r^2 \rangle^{1/2} \simeq 0.34$ fm \cite{Ia73} and $\simeq 0.49$ fm 
\cite{Bi04}, whereas $\alpha$ is related to the coupling of the axial meson. 
Finally, the contributions of the core and the meson cloud to 
$\langle r^2 \rangle_A$ are $-1.71$ and $10.94$ GeV$^{-2}$ for IJL, and 
$0.31$ and $9.64$ GeV$^{-2}$ for BI. Therefore, both for IJL and BI the dominant 
contribution to the axial radius of the nucleon comes from the meson cloud. 

We note, that the negative sign of the contribution of the quark core to the 
nucleon axial radius for the IJL parametrization is related to the change in 
sign of the axial form factor at $Q^2=m_A^2 / (\alpha-1)= 2.65$ GeV$^2$ and the 
occurrence of a pole in the physical region at $t_0=1/\gamma=4$ GeV$^2$, which 
indicates that BI is the preferred parametrization. This is not surprising, 
since the IJL and BI parameters were determined in a fit to experimental data 
available in 1973 and 2004, respectively. 

\section{Time-like region}

The extension of the axial form factor of the nucleon in the two-component model 
to the time-like region can be done by analytic continuation, just as for the case 
of the electromagnetic form factors \cite{Wa05,Bi04}:  
(i) the kinematical variable $Q^2$ is changed into $Q^2 \to -t$, 
(ii) a complex phase $e^{i\delta}$ is introduced into the 
intrinsic form factor of Eq.~(\ref{eq:eq1}), similar to Refs.~\cite{Wa05,Bi04}, and 
(iii) the vector-meson dominance term corresponding to the exchange of an axial 
meson has to be modified in order to take into account the considerable width of 
the axial meson. Here it has been substituted by a Breit-Wigner formula with 
$\Gamma_A=400$ MeV. These modifications lead to the following expression for the 
axial form factor in the time-like region
\begin{equation}
G_A(t)=G_A(0)g(t)\left [1-\alpha +\alpha \frac{m_A^2\left(m_A^2-t+im_A\Gamma_A\right)}{\left(m_A^2-t\right)^2+(m_A\Gamma_A)^2} \right] ~, 
\label{time}
\end{equation}
with
\begin{equation}
g(t)=\left (1-e^{i\delta} \gamma t\right )^{-2} ~. 
\label{gtime}
\end{equation}

Once the parameter $\alpha$ has been determined from the space-like data, the time-like 
behavior of nucleon axial form factor can be calculated using Eqs.~(\ref{time},\ref{gtime}). In 
Fig.~\ref{fig:2}, we show the axial form factor in the space-like ($t<0$) and time-like 
($t>0$) regions for the two-component model obtained from Eq.~(\ref{time}) with 
$\alpha=1.57$, $\gamma=0.25$ GeV$^{-2}$ and $\delta=0.925$ \cite{Ia73,Wa05} (dashed line), 
and $\alpha = 0.95$, $\gamma = 0.515$ GeV$^{-2}$ and $\delta=0.397$ \cite{Bi04} (solid line), 
the dipole form of Eq.~(\ref{eq:dipole}) with $M_A = 1.069$ GeV (dotted line) and the 
experimental data used in the fit of the axial form factor in the space-like region.  

Even though the different parametrizations of the axial form factor coincide in the 
range of $0< Q^2 < 1$ GeV$^2$, outside this range they show large and important 
differences. The position and the shape of the peak in the time-like region is determined 
by the values of $\gamma$ and $\delta$ in the intrinsic form factor. It is interesting to 
note that, outside the region of the peak, the magnitude of the axial form factor is 
significantly higher in the time-like region than in the space-like region. Moreover, 
contrary to the other calculations, the IJL parametrization \cite{Ia73} predicts a zero 
at $Q^2=2.65$ GeV$^2$ in the space-like region. 

\section{Conclusions}

In conclusion, we have proposed a new parametrization of the existing space-like data 
for the axial nucleon form factor by means of a two-component model of the nucleon. 
The physical interpretation of this model corresponds to a compact core surrounded by 
a meson cloud. This parametrization satisfies the analytical properties of the form 
factors and can be extended to the whole kinematical region. The axial form factor 
of the two-component model displays a behavior similar to that of the dipole parametrization 
in the space-like region up to $Q^2=1$ GeV$^2$, whereas outside this region the behavior 
is quite different: IJL predicts a zero around $Q^2=2.65$ GeV$^2$, 
whereas the dipole and BI do not show a change of sign. 

It is important to note that the values of the axial form factor extracted from the 
experimental data in the space-like region are model-dependent, whereas in the time-like 
region there is no experimental information available. A possible way to access the axial 
form factor in the time-like region and in the unphysical region (below the reaction threshold) 
has been suggested through the reactions $N \bar p \to \gamma^* N \pi$ and the crossed channels 
\cite{Re65,Du95,Ad07}. The cross section related to these processes is large and such experiments 
may be performed in future colliders, such as FAIR (Germany), BES3 (China), DANAE (Italy). 
Such experiments also seem to be encouraged by our finding of a non-negligible 
time-like axial form factor, at least up to a few GeV$^2$, as shown in Fig.~\ref{fig:2}. 

We have also discussed the importance of accurate knowledge of the axial form factor in order 
to be able to extract good data on the strange form factors in parity-violating experiments. 
Possible improvements of the present analysis, which will be required in the event 
of new and more precise data, can be foreseen in two directions. First, since the axial 
meson $a_1$ has a large decay width, even larger than that of the $\rho$ meson, 
the corresponding propagator has to be modified to a more complicated form, 
similar to what was done for the $\rho$ meson \cite{Ia73}. Secondly, one may consider the 
contribution of other axial mesons with higher masses. 
A similar study can be applied to the pseudoscalar nucleon form factors. 

\section{Acknowledgments}

M. P. Rekalo is acknowledged for enlightening discussions and ideas. 
We thank F. Iachello for his important remarks and for his interest in this work. 
Thanks are due to U. Meissner for sending the data in tabulated form. 
This work was supported in part by the Slovak Grant Agency for Sciences VEGA under Grant N. 2/4099/26 (C.A.), by CONACYT, Mexico (R.B.) and by INFN (E.S.).

{}

\clearpage\newpage

\clearpage\newpage

\end{document}